\documentclass[aps,preprint]{revtex4}
\usepackage{amssymb}

\usepackage{amsmath}
\usepackage{graphicx}
\usepackage{bm}
\usepackage{mathrsfs}
\usepackage{float}
\usepackage{color}

\begin{document}

\title{Short wavelength {quantum electrodynamical} correction to cold plasma--wave propagation}
\author{J.\ Lundin, G.\ Brodin and M.\ Marklund}
\affiliation{Department of Physics, Ume{\aa} University, SE--90187 Ume{\aa}, Sweden}
\date{\today}

\begin{abstract}
The effect of short wavelength quantum electrodynamic (QED) correction on plasma--wave propagation is investigated. The effect on plasma oscillations and on electromagnetic waves in an unmagnetized as well as a magnetized plasma is investigated. The effects of the short wavelength QED corrections are most evident for plasma oscillations and for extraordinary modes. In particular, the QED correction allow plasma oscillations to propagate, and the extra-ordinary mode looses its stop band. The significance of our results is discussed.
\end{abstract}

\maketitle

\section{Introduction}

The non--linear properties of the quantum vacuum has become increasingly interesting (see e.g. Refs.\ \cite{DiPiazza1,Tiggelen,Zavattini,Rabadan,Blaschke,DiPiazza2,DiPiazza3} for some recent examples), in particular due to the rapidly growing power of present day laser systems \cite{Tajima,Mourou}. It is expected that already the next generation laser systems will reach intensities where quantum electrodynamic (QED) effects, such as electron--positron pair creation and elastic photon--photon scattering, will be directly observable \cite{Ringwald, Lundstrom}. These effects may even play an important role in future laser--plasma experiments \cite{Marklund}, e.g., in laser self focusing \cite{Bulanov, Shorokhov} where laser pulse compression could give rise to field strengths close to the critical field strength, $E_{\text{crit}}= m^2c^3/\hbar e \approx 10^{18}\ \mathrm{V/m}$, where $m$ is the electron rest mass, $c$ is the speed of light in vacuum, $\hbar$ is Planck's constant divided by $2\pi$, and $e$ is the magnitude of the electron charge. Furthermore, in astrophysical environments highly energetic phenomena may give rise to parameter ranges in which QED can be influential. One such example is pulsar magnetospheres \cite{Beskin} and magnetars \cite{Kouveliotou}. In the latter the magnetic fields can reach values above $10^{14}\,\mathrm{G}$, thus giving equivalent electric fields above the Schwinger limit. Thus, in such highly magnetized environments any plasma would be one--dimensional and local gamma-ray sources may be prolific. 

These QED effects arise due to the intense electromagnetic field interacting with the quantum vacuum. By contrast, in this paper we have chosen to look for linear QED effects in plasma wave propagation when the intensity is low. Instead, the frequency is high such that short wavelength corrections may be important. The study is carried out for both plasma oscillations and for electromagnetic waves in an unmagnetized as well as a magnetized plasma. For high plasma densities or high frequency waves, it is found that the short wavelength QED corrections can give appreciable corrections to plasma oscillations and the propagation of extraordinary electromagnetic modes. We discuss possible applications of the results.

\section{Basic equations}

An effective theory for photon--photon scattering can be formulated trough the Heisenberg--Euler Lagrangian density \cite{Heisenberg-Euler, Schwinger}, which describes a vacuum perturbed with a constant electromagnetic field. It is valid for field strengths much lower than the critical field strength and for oscillation length scales much longer than the Compton wavelength. By adding derivative correction terms to the Lagrangian density, the effect of rapid field variations can be taken into account \cite{Mamaev-1981}. However, electron-positron pair creation is not included in this model, and thus the frequencies must still be much lower than the Compton frequency \cite{Marklund,probing} $\omega _{e} = mc^2/\hbar$, and the dispersive and diffractive effects must be small \cite{Rozanov-1998}. The Lagrangian density is 
\begin{eqnarray}
	&&\!\!\!\!\!\!\mathcal{L}=\mathcal{L}_{0}+\mathcal{L}_{HE}+\mathcal{L}_{D} \notag \\ &&\!\!\!\!=\frac{\epsilon _{0}}{4}F_{ab}F^{ab}+\frac{\epsilon _{0}^{2}\kappa }{16}\left[ 4\left( F_{ab}F^{ab}\right) ^{2}+7\left( F_{ab}\hat{F}^{ab}\right) ^{2}\right] +\sigma \epsilon _{0}\left[ \left( \partial_{a}F^{ab}\right) \left( \partial _{c}F_{b}^{c}\right) -F_{ab}\Box
	F^{ab}\right]
\end{eqnarray}
where $\mathcal{L}_{0}$ is the classical Lagrangian density, while $\mathcal{L}_{HE}$ represents the Heisenberg--Euler correction due to first order non--linear QED effects, $\mathcal{L}_{D}$ is the derivative correction, $F^{ab}$ is the electromagnetic field tensor, $\widehat{F}^{ab}=\epsilon ^{abcd}F_{cd}/2$ and $\Box =\partial_{a}\partial ^{a}$. The parameter $\kappa =2\alpha ^{2}\hbar^{3}/45m^{4}c^{5}$ gives the non--linear coupling, the parameter $\sigma =(2/15)\alpha c^{2}/\omega _{e}^{2}$ gives the dispersive effects of the polarized vacuum, and $\alpha = e^2/4\pi \hbar c\epsilon_0$ is the fine structure constant, where $\epsilon_0$ is the free space permittivity. By introducing the four potential $A^{b}$ such that $F_{ab}=\partial _{a}A_{b}-\partial _{b}A_{a}$, we obtain the field equations from the Euler--Lagrange equations $\partial _{b}\left[\partial \mathcal{L}/\partial F_{ab}\right] =\mu _{0}j^{a}$ \cite{Rozanov-1998,Shukla-2004c}, 
\begin{equation*}
\left( 1+2\sigma \Box \right) \partial _{a}F^{ab}=2\epsilon_{0}\kappa \partial _{a}\left[ \left( F_{cd}F^{cd}\right) F^{ab}+\tfrac{7}{4} \left( F_{cd}\widehat{F}^{cd}\right) \widehat{F}^{ab}\right] + \mu_{0}j^{b},
\end{equation*}
where $j^a$ is the four--current and $\mu_{0}$ is the free space permeability.

We further assume that a high frequency low amplitude field is considered, such that the non--linear coupling can be neglected compared to the dispersive effects, i.e. the terms containing $\kappa $ is negligible compared to the terms containing $\sigma $. {To be more precise, we require that $\epsilon_0\kappa|F_{ab}|^2 \ll \sigma\omega^2/c^2$, i.e.\ $|F_{ab}|/E_{\mathrm{crit}} \ll \omega/\omega_e$, where $\omega$ is the frequency with which the field typically changes.}
The corresponding Maxwell equations resulting from the derivative corrected field equation then become 
\begin{subequations}
\begin{eqnarray}
\left[ 1+2\sigma \left( \nabla ^{2}-\frac{1}{c^{2}}\frac{\partial ^{2}}{ \partial t^{2}}\right) \right]\nabla \cdot \mathbf{E} &=& \frac{\rho}{\epsilon_0}, \label{eq:E1a} \\ 
\left[ 1+2\sigma \left( \nabla ^{2}-\frac{1}{c^{2}}\frac{\partial ^{2}}{\partial t^{2}}\right) \right] \left( -\frac{1}{c^{2}}\frac{\partial \mathbf{E}}{\partial t}+\nabla \times \mathbf{B}\right) &=&\mu _{0}\mathbf{j}, \label{eq:E1d}
\end{eqnarray}
where $\rho $ is the total charge density and $\mathbf{j}$ is the current density, while the source-free Maxwell equations read $\nabla \cdot \mathbf{B} =0$ and 
\begin{eqnarray} 
\nabla \times \mathbf{E} &=&-\frac{\partial \mathbf{B}}{\partial t}. \label{eq:E1c} 
\end{eqnarray}
\end{subequations}
The fluid continuity and force equations become 
\begin{eqnarray}
\frac{\partial n_{e}}{\partial t}+\nabla\cdot \left( n_{e}\mathbf{V}_{e}\right) &=&0,\label{eq:E2} \\ 
\left(  \frac{\partial}{\partial t} + \mathbf{V}_{e}\cdot \nabla \right) \mathbf{V}_{e} &=&-\frac{e}{m}\left(\mathbf{E} + \mathbf{V}_{e}\times \mathbf{B} \right),\label{eq:E3}
\end{eqnarray}
where $n_{e}$ is the electron density and $\mathbf{V}_{e}$ is the electron fluid velocity. In order to clearly distinguish the QED effects, thermal effects will be neglected throughout this paper except in the discussion of plasma oscillations where they will be compared to the QED corrections. Since a finite temperature introduces dispersive effects for plasma oscillations, thermal plasma dynamics will compete with the QED corrections (See the Appendix). We note that thermal effects may be important in comparison to the QED corrections for other wave modes, but this is a topic for future research.

Since we focus on high frequency phenomena, the ion-motion is omitted. The charge density and the current density can then be written as 
\begin{eqnarray}
\rho &=& -e(n_e - n_{i0}),\label{eq:E4} \\
\mathbf{j} &=&-en_{e}\mathbf{V}_{e},\label{eq:E5}
\end{eqnarray}
where $n_{i0}$ is the constant background ion density.


\section{Plasma oscillations}
\begin{figure}[H]
\includegraphics[width=0.5\columnwidth]{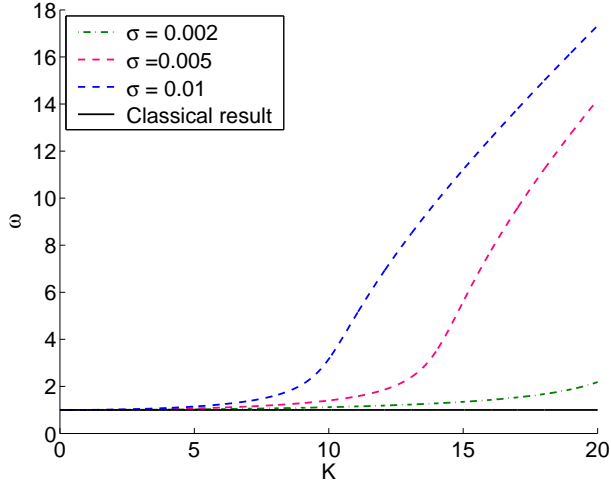}
\caption{Dispersion diagram for a plasma oscillations in a cold plasma for different values of $\bar{\sigma}$. The solid line is the classical dispersion diagram.}
\label{fig:Electrostatic}
\end{figure}

Because of the simple classical dispersion relation for plasma oscillations, $\omega ^{2}=\omega _{p}^{2}$ for a cold plasma, the dispersive effects due to the QED derivative correction are easy to recognize. Below, we neglect thermal effects in our calculations. A complementary discussion where thermal effects are included in plasma oscillations is found in the Appendix. Linearizing Eqs. (\ref{eq:E1a}) and (\ref{eq:E2})--(\ref{eq:E4}) and Fourier decomposing gives us the dispersion relation
\begin{equation*}
\left[ 1-2\sigma \left( k^{2}-\frac{\omega ^{2}}{c^{2}}\right) \right] \omega ^{2}-\omega _{p}^{2}=0,
\end{equation*}
where $\omega _{p}\equiv \sqrt{e^{2}n_{0}/\epsilon _{0}m}$ is the plasma frequency. We normalize the parameters according to 
\begin{equation*}
\bar{\sigma }=\frac{2\omega _{p}^{2}}{c^{2}}\sigma ,\,\,\,\,\,\,\,\, \bar{K}^{2}=\frac{k^{2}c^{2}}{\omega _{p}^{2}},\,\,\,\,\,\,\,\, \bar{\omega }=\frac{\omega }{\omega _{p}}.
\end{equation*}
The dispersion relation is then written as 
\begin{equation*}
\left[ 1-\bar{\sigma}\left( \bar{K}^{2}-\bar{\omega}^{2}\right) \right] \bar{\omega}^{2}-1=0.
\end{equation*}
The deviation from the classical result is illustrated in Fig.\ \ref{fig:Electrostatic}. It is seen that the for small wave numbers, $\bar{K}$, the solution follows the classical one (solid line), but the deviation becomes stronger as $\bar{K}$ increases. It is interesting to note that for sufficiently large $\bar{K}$, the phase velocity approaches that of the speed of light in vacuum, whereas in the classical case there would be no propagation at all. However, within this model the QED corrections must remain small such that the condition $\bar{\sigma}(\bar{K}^{2}-\bar{\omega}^{2})\ll 1$ holds.

\section{Electromagnetic wave in an unmagnetized plasma}
\begin{figure}[H]
\includegraphics[width=0.5\columnwidth]{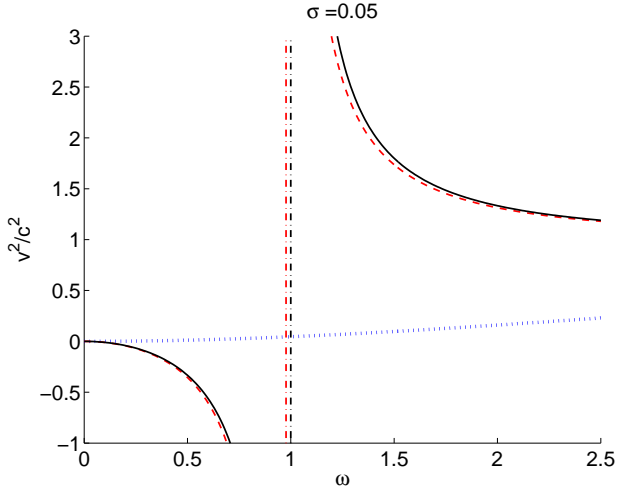}
\caption{The $\bar{v}^{2}=v_{\phi}^{2}/c^{2}$ vs. $\bar{\omega}$ dispersion diagram for an electromagnetic wave in an unmagnetized plasma. The dotted line is a non--physical solution to (\ref{eq:emdisp}). The solid line is the classical dispersion diagram.}
\label{fig:Electromagnetic}
\end{figure}

Linearizing Eqs. (\ref{eq:E1d}), (\ref{eq:E1c}), (\ref{eq:E3}) and (\ref{eq:E5}) and Fourier analysing, the normalized dispersion relation is
\begin{eqnarray}\label{eq:emdisp}
	\left[ 1-\bar{\sigma}\left( \bar{K}^{2}-\bar{\omega}^{2}\right) \right] \left( \bar{\omega}^{2}-\bar{K}^{2}\right) -1=0.
\end{eqnarray}
Using the normalized phase velocity $\bar{v}$, where $\bar{v}=\bar{\omega}/\bar{K}=v_{\phi }/c$, where $v_{\phi }$ is the phase velocity, we can write the dispersion relation as 
\begin{equation*}
\left[ 1-\left( \bar{\sigma}\bar{\omega}^{2}+1\right) \bar{\omega}^{2}\right] \bar{v}^{4}+\left(1+2\bar{\sigma}\bar{\omega}^{2}\right)\bar{\omega}^{2}\bar{v}^2-\bar{\sigma}\bar{\omega}^{4} {=0},
\end{equation*}
which has two distinct solutions. From Fig.\ \ref{fig:Electromagnetic}, it is clearly seen that one of the solutions gives the classical limit when $\bar{\sigma} \rightarrow 0$. For small $\bar{\omega }$, the other root can approximately be written as $\bar{v}^2\approx \bar{\sigma}\bar{\omega}^{2}$ to the lowest non--vanishing order. But $\bar{v}^2=\bar{\omega}^{2}/\bar{K}^{2}$, which implies that $\bar{K}^{2}\approx 1/\bar{\sigma}$, and this violates the condition $\bar{\sigma }(\bar{K}^{2}-\bar{\omega }^{2})\ll 1$, i.e., the dispersive effects are no longer small. For this reason, the non--classical root to the dispersion relation is found to be a non--physical one.

The effect of the corrections due to a rapidly varying field is to dislocate the cut--off frequency to a slightly lower frequency. The shift in frequency is very small and does not become significant until $\bar{\sigma}$ is much larger than allowed in our model. Thus, we can say that within this model, an electromagnetic wave in a unmagnetized plasma is virtually unaffected by the short wavelength QED correction.


\section{Electromagnetic wave, $\mathbf{B}\left\|\mathbf{k}\right.$}
Next we look at wave propagation parallel to an external magnetic field, $\mathbf{B}_{0}=B_{0}\hat{\mathbf{z}}$, where the electric field is circularly polarized,  
\begin{equation*}
\mathbf{E}_{1}=\tilde{E}_{1}(\hat{\mathbf{x}}\pm i\hat{\mathbf{y}})\exp(ikz-i\omega t).
\end{equation*}Linearizing Eqs. (\ref{eq:E1d}), (\ref{eq:E1c}), (\ref{eq:E3}) and (\ref{eq:E5}) we find the normalized dispersion relation to be 
\begin{equation*}
1=\left( \bar{\omega}\pm \bar{\Omega}\right) \left[ 1-\bar{\sigma}\left( \bar{K}^{2}-\bar{\omega}^{2}\right) \right] \left( \bar{\omega}-\bar{K}^{2}/\bar{\omega}\right) ,
\end{equation*}
where $\bar{\Omega}=\omega _{c}/\omega _{p}$, $\omega _{c}\equiv -eB_{0}/m$ is the electron gyro frequency and the +(--) sign means a right hand circularly polarized wave (left hand circularly polarized wave), also called \textit{R}--wave (\textit{L}--wave). The dispersion relation can also be expressed in terms of the normalized phase velocity $\bar{v}$ according to 
\begin{equation*}
\left[ 1-\left( \bar{\sigma}\bar{\omega}^{2}+1\right) \left( \bar{\omega}\pm \bar{\Omega}\right) \bar{\omega}\right] \bar{v}^{4}+\left( 1+2\bar{\sigma} \bar{\omega}^{2}\right) \left( \bar{\omega}\pm \bar{\Omega}\right) \bar{\omega}\bar{v}^2-\bar{\sigma}\bar{\omega}^{3}\left( \bar{\omega}\pm \bar{\Omega}\right) =0.
\end{equation*}

\begin{figure}[H]
\includegraphics[width=0.5\columnwidth]{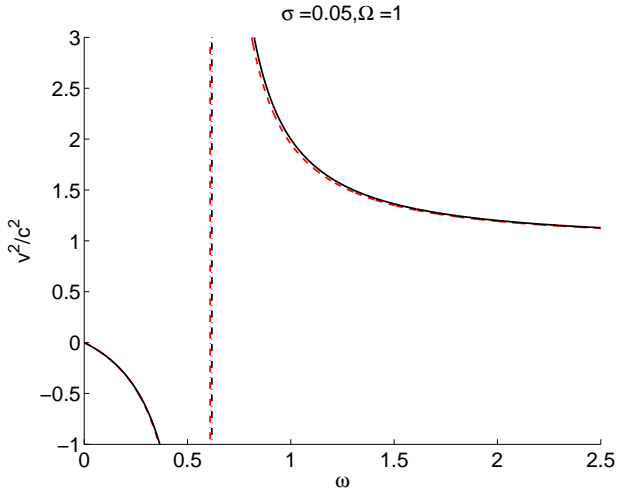}
\caption{The $\bar{v}^{2}=v_{\phi}^{2}/c^{2}$ vs. $\bar{\omega}$ dispersion diagram for an $L$--wave. The solid line is the classical dispersion diagram.}
\label{fig:Lwave}
\end{figure}

\begin{figure}[H]
\includegraphics[width=0.5\columnwidth]{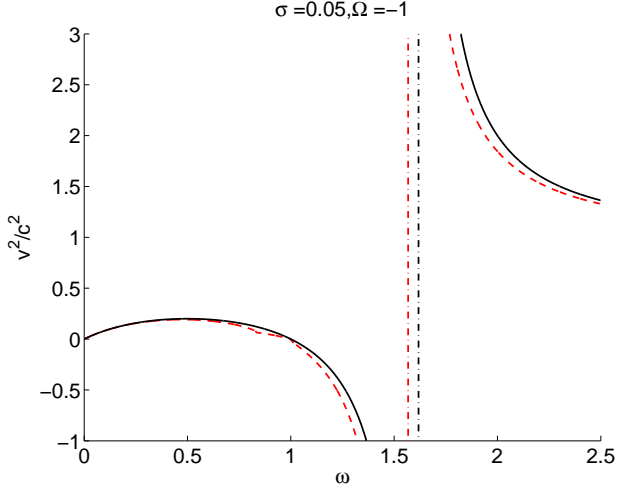}
\caption{The $\bar{v}^{2}=v_{\phi}^{2}/c^{2}$ vs. $\bar{\omega}$ dispersion diagram for an $R$--wave. The solid line is the classical dispersion diagram.}
\label{fig:Rwave}
\end{figure}

In the limit of no magnetic field ($\bar{\Omega}\rightarrow 0$), the dispersion relation simply becomes that of an electromagnetic wave propagating in a unmagnetized plasma, (\ref{eq:emdisp}), which had a non--physical root. The corresponding non--physical root has been disregarded in Fig.\ \ref{fig:Lwave} and Fig.\ \ref{fig:Rwave}.

The effect of the corrections due to a rapidly varying field is again to dislocate the cut--off frequency of the \textit{R}-wave and the \textit{L}--wave to a slightly lower frequency. Even though the effect is more pronounced for the \textit{R}-wave than for the \textit{L}-wave, see Fig.\ \ref{fig:Lwave} and Fig.\ \ref{fig:Rwave}, it is still very small. 



\section{Electromagnetic wave, $\mathbf{B}\bot\mathbf{k}$ ($X$--wave)}
\begin{figure}[H]
\includegraphics[width=0.5\columnwidth]{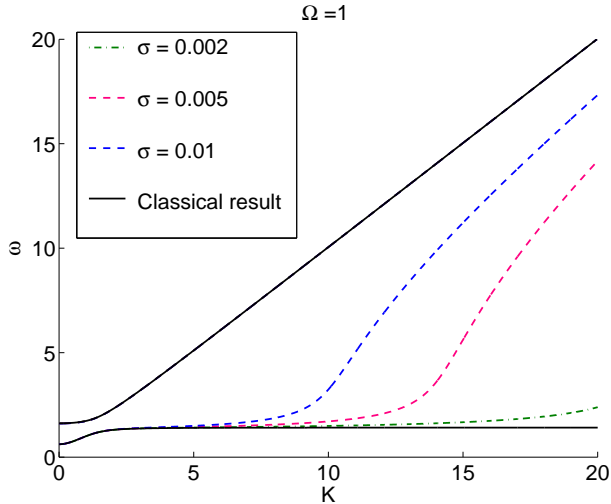}
\caption{Dispersion diagram for an $X$--wave for different values of $\bar{\sigma}$. The solid line is the classical dispersion diagram.}
\label{fig:Bortodisp}
\end{figure}

We now consider an electromagnetic wave propagating orthogonal to the magnetic field of a magnetized plasma, such that the electric field oscillates orthogonal to the external magnetic field ($X$--wave). Linearizing Eqs (\ref{eq:E1d}), (\ref{eq:E1c}), (\ref{eq:E3}) and (\ref{eq:E5}) we find the normalized dispersion relation to be
\begin{eqnarray}\label{eq:dispXwave}
\left\{ 1-\bar{\omega}^{2}\left[ 1-\bar{\sigma}\left( \bar{K}^{2}-\bar{\omega}^{2}\right) \right] \right\} \left\{ 1+\left( \bar{K}^{2}-\bar{\omega}^{2}\right) \left[ 1-\bar{\sigma}\left( \bar{K}^{2}-\bar{\omega}^{2}\right) \right] \right\} + &&  \notag  \label{eq:dispBorto} \\+\bar{\Omega}^{2}\left( \bar{K}^{2}-\bar{\omega}^{2}\right) \left[ 1-\bar{\sigma}\left( \bar{K}^{2}-\bar{\omega}^{2}\right) \right] ^{2} &=&0.
\end{eqnarray}
In the limit of no magnetic field ($\bar{\Omega} \rightarrow 0$), the dispersion relation becomes a product of that for plasma oscillations and that for an electromagnetic wave in a unmagnetized plasma. The corresponding non--physical root of (\ref{eq:dispXwave}) has been disregarded in Fig. \ref{fig:Bortodisp}.

The solutions of (\ref{eq:dispBorto}) are easily identified with the two classical ones (solid lines) in Fig.\ \ref{fig:Bortodisp}. As can be seen, the high frequency branch does not significantly deviate from the classical one. However, the deviation is more pronounced for the low frequency branch. While it is not explicitly shown in Fig.\ \ref{fig:Bortodisp}, it should be noted that the QED induced effects depend only weakly on the external magnetic field. In Fig.\ \ref{fig:Bortodisp}, this solution has been plotted for different values of the normalized dispersive parameter, $\bar{\sigma}$. It is interesting to note that the short wavelength correction removes the stop band of the $X$--wave.

\section{Summary and discussion}
In this paper we have investigated how the short wavelength QED correction affects plasma wave propagation in both a unmagnetized and a magnetized plasma. In order to concentrate on the effects associated with the QED corrections, we have chosen to consider the simple case of a cold plasma. Furthermore, the field amplitude is assumed to be small such that the nonlinear effects can be neglected. The dispersive effects due to the short wavelength QED correction is found to be small but well pronounced for plasma oscillations, Fig.\ \ref{fig:Electrostatic}, and $X$--waves, Fig.\ \ref{fig:Bortodisp}, for sufficiently large wavenumbers, whereas the effect is less pronounced for parallel propagating $R$-- and $L$--waves, Fig.\ \ref{fig:Rwave} and Fig.\ \ref{fig:Lwave}, and electromagnetic waves in a unmagnetized plasma, Fig.\ \ref{fig:Electromagnetic}. We know that the condition $\bar{\sigma}(\bar{K}^{2}-\bar{\omega}^{2})\ll 1$ and $\omega \ll \omega _{e}$ must be satisfied, for our model to be applicable, with $\bar{\sigma}\approx (1/500)(\omega _{p}^{2}/\omega _{e}^{2})$. For all of the waves considered above, except for $R$--waves, we learn from the dispersion diagrams that $\omega \geq \omega _{p}$ always, at least for sufficiently large wavenumbers $\bar{K}$. Thus, for sufficiently dense plasmas, we will reach parameter ranges where the dispersive QED effects become large. The highest plasma densities on earth are found in laser fusion experiments where the density can reach values as high as $n_{0}\approx 10^{28}\,\text{cm}^{-3}$ \cite{icf}, giving $\omega _{p}\approx 0.01\omega_{e}$. For such plasma densities, the dispersive parameter becomes, $\bar{\sigma}\approx 2\times 10^{-7}$. Detection of these QED corrections must therefore be done in the extreme short wave limit $\bar{K}\gg 1$ or with very high precision. The phase shift of a laser beam can be measured with extremely high precision through interferometry \cite{interferometry}. Also, by using a experimental setup consisting of a standing wave in a plasma, the phase shift at each pass through the plasma can be accumulated, thus enhancing the total phase shift to be measured. However, to single out the dispersive QED effects from the general dispersive effects of a plasma requires detailed knowledge of the plasma parameters for such an experiment to be conclusive. In this respect the conditions in laser fusion experiments are not suitable, due to the rapid change of plasma density. For lower plasma densities, a fairly high stability of the plasma conditions could compensate for the even smaller value of $\bar{\sigma}$ to make a high precision experiment possible. In an experiment, the plasma parameters may either be determined with independent measuring techniques, but it might also be possible to extract information about the plasma by measuring the phase shift for different frequencies or polarizations. In this case, thermal effects may be of importance (see the Appendix), and the inclusion of such is therefore of interest for future research. How to extract the phase information and how to construct a scheme to detect the QED short wave corrections is nontrivial and requires extensive work.

\acknowledgments

This research was supported by the Swedish Research Council Contract No. 621-2004-3217.

\section{Appendix}

In order to highlight the QED effects we have considered the simple case of a cold plasma in the manuscript. However, since the QED effects in most cases are small corrections, typically even very low temperatures give a comparatively large contribution to the dispersion relations studied, except for wave modes that have density perturbations that are identically zero. To illustrate the relative contribution from QED and thermal effects, we reconsider the case of plasma oscillations by adding a pressure term {$-\nabla P/mn_e$} to the right hand side of Eq. (\ref{eq:E3}). Proceeding as in section III, we linearize the pressure term according to $-\nabla P/mn_e=-v_{t}^{2}\nabla n_{e}/n_0$ (where deviation from isothermal behavior is incorporated in the definition of $v_{t}^{2}$), and the dispersion relation is then modified to 
\begin{equation*}
\left[ 1+2\sigma \left(\frac{\omega ^{2}}{c^{2}} - k^{2}\right) \right]\left(\omega ^{2}-k^{2}v_{t}^{2}\right)=\omega_{p}^{2},
\end{equation*}
which (given  $\sigma (\omega^2/c^2-k^{2})\ll 1$) can be approximated by 
\begin{equation*}
\omega ^{2}=\omega_p^2\left[ 1-2\sigma \left(\frac{\omega ^{2}}{c^{2}} - k^{2}\right) \right]+k^{2}v_{t}^{2}.
\end{equation*}
In the regime where Landau damping is small, $kv_{t}\ll \omega _{p}$, we can
omit the therm proportional to both QED and thermal corrections and write to
first order 
\begin{equation*}
\omega ^{2}=\omega _{p}^{2}\left( 1-\frac{2\sigma \omega _{p}^{2}}{c^{2}}%
\right) +k^{2}\left( v_{t}^{2}+2\sigma\omega_p^2 \right) .
\end{equation*}
Thus the QED effects give a relative frequency shift $2\sigma \omega
_{p}^{2}/c^{2}$ of the plasma frequency, and a modification of the group
velocity of the Langmuir waves  proportional to $2\sigma\omega_p^2 /v_{t}^{2}$. We
note that the group velocity shift is a small parameter for any reasonable
plasma temperature. This suggests that in order to avoid confusion with
small temperature fluctuations in a possible detection scheme of the QED
effects, it might be preferable to look for a small shift in plasma
frequency, rather than the group velocity shift.

\end{document}